\documentclass[12pt]{article}
\def\slash#1{\setbox0=\hbox{$#1$}#1\hskip-\wd0\hbox to\wd0{\hss\sl/\/\hss}}
\usepackage{epsf, cite}
\usepackage{epsfig}
\usepackage[all]{xy}
\usepackage{amsfonts}
\usepackage{latexsym}
\usepackage{amsmath}
\usepackage{amssymb}
\usepackage{cite}

\usepackage{epsf, cite}

\makeatletter
\renewcommand\section{\@startsection {section}{1}{\z@}%
                                   {-3.5ex \@plus -1ex \@minus -.2ex}%nn
                                   {2.3ex \@plus.2ex}%
                                   {\normalfont\large\bfseries}}
\renewcommand\subsection{\@startsection{subsection}{2}{\z@}%
                                     {-3.25ex\@plus -1ex \@minus -.2ex}%
                                     {1.5ex \@plus .2ex}%
                                     {\normalfont\bfseries}}
\makeatother

\def\lbldef#1#2{\expandafter\gdef\csname #1\endcsname {#2}}

\def\href#1#2{#2}
\def\beq{\begin{equation}}
\def\eeq{\end{equation}}
\def    \bea    {\begin{eqnarray}}
\def    \eea    {\end{eqnarray}}
\def \d {\partial}

\begin{document}
\pagestyle{plain}
\begin{titlepage}

\begin{center}

\hfill{QMUL-PH-2008-05} \\

\vskip 1cm

{{\Large \bf Aspects of Multiple Membranes}} \\

\vskip 1.25cm {David S. Berman\footnote{email: D.S.Berman@qmul.ac.uk},
Laura C. Tadrowski\footnote{email: L.C.Tadrowski@qmul.ac.uk} } and
Daniel C. Thompson\footnote{email: D.C.Thompson@qmul.ac.uk}
\\
{\vskip 0.2cm
Queen Mary College, University of London,\\
Department of Physics,\\
Mile End Road,\\
London, E1 4NS, England\\
}

\end{center}
\vskip 1 cm

\begin{abstract}
\baselineskip=18pt\
This paper examines various aspects of the recently proposed theory of
coincident membranes by Bagger and Lambert. These include the 
properties of open membranes
and the resulting boundary theory with an interpretation in terms of
the fivebrane and marginal supersymmetric deformations of the
interactions with the relation to the holographic dual.

\end{abstract}

\end{titlepage}

\pagestyle{plain}

\baselineskip=19pt

\section{Introduction}

Two of the most outstanding problems in M-theory are the understanding
of the world volume theories of coincident membranes and coincident
fivebranes \cite{Berman:2007bv}. In a fascinating recent development Bagger
and Lambert \cite{Bagger:2006sk, Bagger:2007jr, Bagger:2007vi} have proposed a Lagrangian description
of coincident interacting membranes which relies on the scalar fields
of the membrane world volume taking values in some non-associative
algebra. (A similar approach was also described by Gustavsson in
\cite{Gustavsson:2007vu}). Remarkably, the theory could be made to possess the ${\cal{N}}=8$
supersymmetry required of a membrane with the introduction of a
{\it{gauge}} connection whose kinetic piece is a twisted version of
Chern-Simons theory. The goal of this paper is to investigate
properties of this theory. The two aspects will both involve breaking
the supersymmetry in interesting ways. The first, through introduction
of a boundary and the second, by directly altering the interaction
potential while preserving some supersymmetry.

First, we will consider the
theory of open membranes i.e. the Bagger Lambert theory with a
boundary. The motivation for this is to see aspects of
fivebrane physics from the point of view of the membrane boundary. If the Bagger Lambert
action is a good description of the membrane then this would be
expected since the fivebrane is the M-theory analogue of a D-brane.

The key issue is to examine the twisted Chern-Simons
theory since this gives rise to dynamical degrees of freedom on the
membrane boundary. In fact we will show
that the membrane boundary is described by a sigma model whose
target space is six dimensional. This is consistent with a 
self-dual string in the fivebrane world volume.

We will then consider possible marginal deformations of the
interactions that preserve some fraction of supersymmetry. This is the
M-theory analogue of the Leigh Strassler deformed ${\cal{N}}=4$
Yang-Mills theory whose supergravity dual was described in \cite{Lunin:2005jy}. We will
examine the relationship between the proposed supersymmetry preserving
deformations of the membrane theory and the deformed supergravity dual
described and investigated in \cite{Berman:2007tf}.

Whether this is the true description of coincident membranes is still
not certain however supersymmetric field theories have proved of
immense interest over the years \cite{Berman:2002kd} and has been shown to possess a rich
and deep theoretical structure. Therefore the arrival of an entirely
new supersymmetric field theory where the fields are nonassociative
demands study.

\section{An effective theory of interacting membranes}

In \cite{Bagger:2006sk} Bagger and Lambert proposed a theory with
$\mathcal{N}=8$ supersymmetry to describe multiple coincident
  membranes.  The novel insight allowing this construction is that the fields take values in a
  nonassociative algebra, denoted here by $\mathcal{A}$.
This nonassociative algebra, also called a three algebra, is endowed
with a bilinear product and a totally antisymmetric three-bracket
instead of the standard commutator found in Lie algebras.
The three bracket or triple product is given by the antisymmetrised associator. For example the associator of three transverse scalars is
\bea
\langle X^I, X^J, X^K \rangle = (X^I\cdot X^J) \cdot X^K - X^I \cdot (X^J \cdot X^K)
\eea
and the three bracket is then
\bea
[ X^I, X^J, X^K  ] = \frac{1}{12} \langle X^{[I}, X^J, X^{K]} \rangle\, .
\eea
One can introduce a basis $\{T^a\}$ of $\mathcal{A}$ satisfying
\newcommand{\f}[2]{f^{#1}_{\phantom{#1} #2}}
\bea
[ T^a, T^b, T^c ] = \f{abc}{d}T^d\, ,
\eea
where the totally antisymmetric structure constants\footnote{We raise and lower algebraic indices with a positive definite trace form metric which, in this paper, we take to be simply $ \delta_{ab}$.} $\f{abcd}{}$ obey the fundamental identity, akin to the Jacobi identity of Lie algebras,  given by
\bea
\label{fundid}
\f{efg}{d} \f{abc}{g} = \f{efa}{g} \f{bcg}{d} + \f{efb}{ g} \f{cag}{ d} + \f{efc}{ g} \f{abg}{ d}\, .
\eea
We remark that at this stage we have not specified the dimension of
the algebra which we shall denote by $n$.
\newcommand{\At}[3]{\tilde{A}^{\phantom{#1} #2}_{#1 \phantom{#2} #3}}
\newcommand{\A}[3]{{A}^{\phantom{#1} #2}_{#1 \phantom{#2} #3}}
\par
To make the supersymmetry algebra close \cite{Bagger:2007jr} it is
necessary to introduce non-propagating fields  $\At{\mu}{b}{a}$, which
gauge the transformation:
\bea
\label{Xtransform}
\delta X^I_a = \Lambda_{cd}\f{cdb}{a} X^I_b \equiv  \tilde \Lambda^b_{\phantom{b}a}  X^I_b\,.
\eea
 The gauge field is antisymmetric as a consequence of the antisymmetry
 of  $\f{cda}{b}$ so the gauge group $G\subseteq SO(n)$. In fact since
 we are only examining the connection, one can only make statements
 about the algebra but it will be assumed that there is a full group structure.

As a consequence of the transformation law (\ref{Xtransform}) the
group $G$ is restricted by insisting that one may write:
\bea
\label{AtildefA}
\At{\mu}{b}{a}= \f{cdb}{a} \A{\mu}{\phantom{.}}{cd}
\eea
for some $n\times n$ matrix valued $\A{\mu}{\phantom{.}}{cd}$ with
$\f{cdb}{a}$ satisfying the fundamental identity which implies $\f{abcd}{}$ must be an invariant four form of the group.
\par
The Lagrangian for the full $\mathcal{N}=8$ theory including these gauge fields is given by
\newcommand{\G}{\Gamma}
\bea
\label{lagrangian}
\nonumber \mathcal{L}& =& -\frac{1}{2}D^\mu X^{aI}D_\mu X^{I}_a +\frac{i}{2}\bar{\Psi}^a\G^\mu D_\mu \Psi_a + \frac{i}{4}\bar{\Psi}_b\G_{IJ}\Psi_aX^I_cX^J_d \f{abcd}{} \\&&-V(X) +\frac{1}{2}\epsilon^{\mu \nu \lambda} \left(\f{abcd}{}\A{\mu ab}{}{} \d_{\nu} \A{\lambda cd}{}{} +\frac{2}{3 }\f{cda}{ g}\f{efgb}{}\A{\mu ab}{}{}\A{\nu cd}{}{}\A{\lambda ef}{}{}\right)\, ,
\eea
with bosonic potential
\bea
V(X) = \frac{1}{12} Tr\left( [X^I,X^J,X^K]^2\right)\, ,
\eea
and supersymmetry transformations
\bea
\label{SUSYtransforms}
\delta X^{I}_a &=& i\bar{\epsilon} \G^{I}\Psi_a\, ,\\
\delta \Psi_a &=& D_\mu X^{I}_a\G^\mu \G^I\epsilon - \frac{1}{6}X^I_bX^J_cX^K_d \f{bcd}{a}\G^{IJK}\epsilon\, ,   \\
\delta \At{\mu}{b}{a} &=& i\bar{\epsilon}\G_\mu\G^IX^I_c \Psi_d\f{cdb}{a}\, ,
\eea
where the covariant derivative acts as $D_\mu X_a = \d_\mu X_a -
\At{\mu}{b}{a}X_b$.  The gauge kinetic term is similar to Chern-Simons
theory but twisted with the structure constants of the algebra.
Obviously, the gauge fields are non-propagating as is required to give
the correct degrees of freedom for supersymmetry. 
\par
In the remains of this paper we shall explore two truncations of this
theory. We shall examine the gauge sector of the theory in closer
detail by switching off the scalars and spinors. This will allow us to investigate the
boundary theory of open coincident membranes.
We shall then restrict our attention to the scalar-spinor sector and
study marginal deformations of the theory that preserve
$\mathcal{N}=2$ supersymmetry.

\section{Twisted Chern-Simons and the self-dual string boundary theory}

When open membranes end on a fivebrane the boundary may be described
by a $\mathcal{N}=(4,4)$ self-dual string theory
(see for example, \cite{Berman:2007bv}).
The self-dual string can be regarded from two perspectives; it can be
viewed simply as a solitonic solution of the
fivebrane world volume equations of motion \cite{Howe:1997ue}  or one
can think of it as the boundary theory of coincident membranes \cite{Townsend:1995af, Strominger:1995ac}.

From the fivebrane perspective one can take a {\it{Maldacena style}} limit and consider the resulting
geometry of the fivebrane to describe the string.
One finds that in this {\it{near horizon}} limit the self-dual
string is described by a fivebrane with $AdS_3\times S^3$ geometry \cite{Berman:2001fs}.

In this section we shall study the self-dual string from the point of
view of the boundary theory of coincident membranes. We will examine the gauge sector in isolation (setting
$X^I=\Psi =0 $) and compare the resulting boundary theory to the
fivebrane description.
\par
To make concrete progress we need to specify a gauge group of the
twisted Chern-Simons theory. This essentially means solving the fundamental identity (\ref{fundid}).
There is only one finite dimensional known solution (though other more
exotic solutions have been discussed in \cite{Gustavsson:2008dy}).
When the dimension of the algebra is four it is known that
\bea
\f{abcd}{} \propto \epsilon^{abcd}
\eea
satisfies the fundamental identity and the associated gauge group is
then $SO(4)$.
There is some evidence in the literature that this $n=4$
algebra, which we denote by $\mathcal{A}_4$, may be the only possible solution to the fundamental identity
\cite{Gustavsson:2008dy , Kawamura:2003cw}.
This restriction to the dimension four algebra is not something that we will find problematic, rather we find that
this will have a natural interpretation in what follows. We will thus restrict ourself to the gauge group $G=SO(4)$.
\par
First we carry out the usual decomposition of $SO(4)\cong SU(2) \times  SU(2)$
 by splitting the $so(4)$ gauge field into self-dual and anti-self-dual parts
using 't Hooft matrices \cite{tHooft:1976fv}.
Let us perform this decomposition in our twisted Chern-Simons theory. We write the gauge field as
 \bea
 \tilde{A} = A^+ + A^- \, ,
 \eea
where
 \bea
\ast A^+ = A^+ \,, \, \ast A^- = - A^-
 \eea
and where $\ast$ denotes Hodge star on the matrix indices i.e.
  \bea
  (\ast A^+)_{ab} =\frac{1}{2} \epsilon^{abcd}A^+_{cd}\,.
\eea
We also have that $\ast^2 = 1$ and that for $\mathcal{A}_4$ the structure constants when viewed as operators have the same action on gauge fields as $\ast$.  Using the relation (\ref{AtildefA}) between $\tilde{A}$ and $A$ we find (omitting indices for convenience) that
\bea
\ast \tilde{A} = A^{+} - A^- = \ast f A = \ast^2 A  = A\, .
\eea
Then the twisted Chern-Simons action becomes
\bea
\nonumber \mathcal{L}_{TCS}  &=&- \frac{1}{4\pi} \epsilon^{\mu\nu\lambda}\left(\f{abcd}{}\A{\mu ab}{}{} \d_{\nu} \A{\lambda cd}{}{} + \frac{2}{3}\f{cda}{ g}\f{efgb}{}\A{\mu ab}{}{}\A{\nu cd}{}{}\A{\lambda ef}{}{}\right)\\
\nonumber  &=& \frac{Tr}{4\pi} \left(A d \tilde A  + \frac{2}{3}A \wedge \tilde A \wedge \tilde A\right)\\
 \nonumber &=&   \frac{Tr}{4\pi} \left(A^+ d A^+  + \frac{2}{3}A^+ \wedge A^+ \wedge A^+\right) - \frac{Tr}{4\pi} \left(A^- d A^-  + \frac{2}{3}A^- \wedge A^- \wedge A^-\right)\\
 &=& \label{decomposedCS} \mathcal{L}_{CS}[A^+] -  \mathcal{L}_{CS}[A^-]\, .
\eea
Implicitly we are working at level $k=1$ and henceforward we shall drop the overall $4\pi$ normalisation of the Lagrangian.  We see that the twisted Chern-Simons action has decomposed into two $SU(2)$ Chern-Simons theories but note that there is an all important relative minus sign between the two\footnote{ Carrying out the same analysis but starting with the un-tilded $A$ leads to an apparent relative minus sign on the $A^-$ cubic term but this is not physically significant since it can be removed by a field redefinition.}.
\par
Let us now consider what happens when our membrane has a boundary.  With a boundary, regular Chern-Simons theory gives rise to a (chiral) Wess-Zumino-Witten theory of propagating boundary degrees of freedom    \cite{Witten:1983ar, Witten:1988hf, Elitzur:1989nr} (see also \cite{Dunne:1998qy} for a helpful review).  The naive first thought is that the boundary theory is a WZW model with a target space $S^3\times S^3$.   However the relative minus sign between the two Chern-Simons terms means we must be much more careful.\par
We shall briefly outline the process to arrive at the WZW model and
see how this minus sign plays a role. We start by considering a
single $SU(2)$ Chern-Simons theory with level $k = 1$.
One writes the three manifold as $M=\Sigma \times \mathbf{R}$ with
boundary $\d M= \d \Sigma \times \mathbf{R}$.
For simplicity we consider only the case where $\Sigma$ is a disc such
that the boundary of $M$ is a cylinder
with angular coordinate $\theta$ and $\mathbf{R}$ identified with
time. Upon doing a gauge variation of the Chern-Simons Lagrangian one picks up a surface term. To eliminate this we choose a suitable boundary condition, for example that the time component of the gauge field $A_0$ vanishes at the boundary. Equipped with this boundary condition one is able to recast the Chern-Simons theory as
  \bea
  \label{intermediateaction}
  S = \int d^3 \sigma \epsilon^{0ij} \mbox{Tr} \left( A_0 F_{ij}   -A_i \d_0 A_j  \right)\, ,
  \eea
  so that in the bulk $A_0$ becomes a Lagrange multiplier enforcing Gauss' law for the spatial components of the gauge field.  Integrating out this Lagrange multiplier places the remaining components of the gauge field to be pure gauge $A_i = g^{-1}\d_i g$.   Writing out the resulting action in terms of the group element one finds
  \beq
 S_{WZW} = - \int_{\d M} d\theta d t \mbox{Tr}\left(g^{-1} \d_\theta g  g^{-1} \d_t g \right) + \frac{1}{3}\int_M  d^3 \sigma \epsilon^{\mu\nu\lambda}\mbox{Tr} \left(g^{-1} \d_\mu g g^{-1}\d_\nu g g^{-1} \d_\lambda g \right)\, .
\eeq
This action is invariant under $g\rightarrow L(\theta)gR(t)$.  $L(\theta)$ corresponds to transformations which do not vanish at $t=\pm \infty$ and therefore states lie in representations of this symmetry.  $R(t)$ is a gauge transformation which must fixed.  Hence the action is actually a chiral Wess-Zumino-Witten theory \cite{Elitzur:1989nr}.
\par
 Let us explicitly calculate the action by parameterising the group element with Euler angles in the form
\bea
g(\theta, t) = e^{i u(\theta, t) \sigma_2}e^{i x(\theta, t) \sigma_3} e^{i v (\theta, t)\sigma_2}\,,
\eea
with
 \bea
 \sigma_2= \left(\begin{array} {cc}
0 &  -i\\ i &  0 \end{array}\right), &
 \sigma_3= \left(\begin{array} {cc}
1 & 0\\ 0 & -1 \end{array}\right).
\eea
We find the kinetic term contributes
\bea
- \mbox{Tr}\left(g^{-1} \d_\theta g  g^{-1} \d_t g \right)  =  2 \left( \d_\theta x \d_t x + \d_\theta u \d_t u+ \d_\theta v \d_t v   + \cos 2x \left(\d_\theta u \d_t v+ \d_\theta v \d_t u     \right)  \right)\, .
\eea
The Wess-Zumino term is simply the pull back of the canonical three form of the group which we calculate as
\bea
\nonumber \chi &=& \frac{1}{3} Tr (g^{-1}dg)^3 \\
\nonumber &=& -4 \sin 2x \, d x\wedge d u \wedge dv\\
&=& d \left( 2\cos 2x\, du\wedge dv\right) \, .
\eea
Altogether we find the WZW action is
\bea
S_{WZW} = 2 \int d \theta dt  \left( \d_\theta x \d_t x + \d_\theta u \d_t u+ \d_\theta v \d_t v  + 2 \cos 2x \, \d_\theta u \d_t v    \right)\, .
\eea
To see this in a more familiar form we do a field redefinition
\bea
u = \frac{1}{2} (\tau + \phi)\, ,  \quad
v =  \frac{1}{2} (\tau - \phi)\, ,
\eea
and find after some trivial trigonometry
\bea
S_{WZW} = \int d \theta d t  \left(\cos^2x\,  \d_\theta \tau \d_t \tau + \sin^2 x\, \d_\theta \phi \d_t \phi +    \d_\theta x \d_tx + \cos 2x \left(\d_\theta \tau \d_t \phi -    \d_\theta \phi \d_t \tau  \right)\right)\,.
\eea
We can now read off the metric, the 2-form and its field strength by picking out the symmetric and antisymmetric terms in this action
\bea
ds^2 &=& \cos^2x\, d \tau ^2 + \sin^2x\, d  \phi^2+  dx^2\, , \\
B &=&  \cos 2x\, d \phi \wedge d \tau \, ,\\
H =dB &=& 2 \sin 2x\, d x \wedge d \phi \wedge d \tau= 4 \sqrt{g} d x \wedge d \phi \wedge d \tau\,.
\eea
The target space is a three-sphere (written in Hopf coordinates) with unit radius and with one unit of normalized flux through the sphere.
\par
For the multiple M2 theory we can read off the complete six dimensional target space metric and field strength by simply taking into account the extra minus sign entering into (\ref{decomposedCS})  to find
\bea
\label{targetmetric1}
ds^2 &=& \cos^2 x\, d \tau ^2 + \sin^2 x\, d  \phi^2+  dx^2   - \cos^2 \tilde x \, d  \tilde \tau ^2 - \sin^2 \tilde x \, d  \tilde \phi^2- d\tilde x^2  \\
H =dB &=& 2 \sin 2x \, d x \wedge d \phi \wedge d \tau  -  2 \sin 2\tilde x\, d \tilde x \wedge d \tilde\phi \wedge d \tilde \tau\,.
\eea
The flux is anti-self-dual i.e. $\ast H = - H$ and the geometry here
can be thought of as a sphere of unit radius cross a sphere of
imaginary (unit) radius.  The sphere of imaginary radius is a space of constant negative
curvature which we identify with $AdS_3$. Indeed, if one analytically continues $\tilde{x}
\rightarrow i \tilde{x}$ so as to absorb the minus signs into the
metric of the target space then the result would be the metric and flux for
$AdS_3\times S^3$
\footnote{This appears to be legitimate locally but one should be very careful in how to view the global nature of the space.}.
This is very satisfactory, from examining the boundary theory we see
the target space is six dimensional with an
anti-self-dual three form flux and a geometry of $AdS_3 \times
S^3$. Comparing this to the fivebrane we see that this
is indeed the description of
the self-dual string in a {\it{Maldacena style}} limit
\cite{Berman:2001fs}. The restriction to $SO(4)$ now makes sense since
it ensures the dimension of the 
boundary theory target space is six i.e. the dimension of the fivebrane.

One may wonder whether one could reinstate $k$, the level, and introduce
a parameter to control the number of membranes in this way. This has one
appealing property that the self-dual string charge will be given by
$k$. However, there are various difficulties with this approach. First,
the supersymmetry requires that we also multiply the matter sector by
$k$ which does not feel appropriate for the interacting membrane
theory. Second, the scaling of radius of the target space $S^3$ with $k$
does not match the scaling found in \cite{Berman:2001fs}.

\section{Marginal deformations}

We now shift our attention back to closed membranes and begin by describing the Bagger Lambert theory in ${\cal{N}}=2$
superspace language. The gauge field is removed and so the
${\cal{N}}=8$ supersymmetry is no longer manifest and only an
$SU(4)\times U(1)\subset SO(8)$ R-symmetry remains.
The fields are redefined using a notation suited to the
$SU(4)\times U(1)$ symmetry as follows:
\begin{eqnarray}
X^I&\rightarrow& \ Z^A\oplus Z_{\bar{A}}\ \ \ \ \ \ \ \in \ \ \ \ \ \ \ \bold{4}(1)\oplus\bar{\bold{4}}(-1),\nonumber\\
\Psi&\rightarrow& \ \psi^A\oplus \psi_{\bar{A}}\ \ \ \ \ \ \ \in \ \ \ \ \ \ \ \bold{4}(-1)\oplus\bar{\bold{4}}(1),\\
\epsilon&\rightarrow&\varepsilon\oplus\varepsilon^{*}\oplus\varepsilon^{AB} \ \ \ \in \ \ \ \bold{1}(-2)\oplus\bar{\bold{1}}(2)\oplus\bold{6}(0)\nonumber.
\end{eqnarray}
The $\mathcal{N}=2$ supervariation in this form is,
\begin{eqnarray}
\delta Z^A &=& i\bar{\varepsilon}\psi^A,\nonumber\\
\delta\psi^A &=& 2\gamma^{\mu}\partial_{\mu}Z^A\varepsilon +
i\kappa_1\epsilon^{ABCD}\big[Z_{\bar{B}},Z_{\bar{C}},Z_{\bar{D}}\big]\varepsilon^{*}+3i\kappa_3\big[Z^A,Z^B,Z_{\bar{B}}\big]
\epsilon \, .
\end{eqnarray}

One may then formulate the theory in terms of $\mathcal{N}=2$ chiral superfields, $\mathcal{Z}^A$, which satisfy
\begin{equation}
\bar{D}\mathcal{Z}^A=0,
\end{equation}
with an expansion
\begin{equation}
\mathcal{Z}^A=Z^{A}(y)+\bar{\theta}^{*}\psi^A(y)+\bar{\theta}^{*}\theta F^A(y),
\end{equation}
where $y^{\mu}=x^{\mu}+i\bar{\theta}\gamma^{\mu}\theta$. The Clifford
algebra has a real basis and conjugation of the complex spinors
$\varepsilon$ and $\theta$ is defined as
$\bar{\theta}=\theta^{*T}\gamma^0$.

The $\mathcal{N}=2$ supersymmetry algebra closes when the following constraints are imposed:
\begin{eqnarray}
\big[Z^A,Z^B,Z_{\bar{B}}\big]&=&0 \, ,\\
\big[Z^A,Z^B,\psi_{\bar{B}}\big]&=&0 \, ,\\
\big[\psi^A,Z^B,Z_{\bar{B}}\big]+\big[Z^A,\psi^B,Z_{\bar{B}}\big]&=&0 \, .
\end{eqnarray}
These follow from the single superspace constraint
\begin{equation}
\big[\mathcal{Z}^A,\mathcal{Z}^B,\mathcal{Z}_{\bar{B}}\big]=0 \,  \label{constraint}
\end{equation}
and result in the vanishing of the third term in the supervariation of the fermion. Hence, in what follows, $\kappa_3$ does not appear in our analysis.
When the dust settles one is left with the following Lagrangian
written in terms of ${\cal{N}}=2$ chiral superfields obeying the
constraint (\ref{constraint}):
\begin{equation}
\mathcal{L}=\frac{1}{2}\int\mathrm{d}^4\theta \
\mathrm{Tr}\big(\mathcal{Z}^A,\mathcal{Z}_{\bar{A}}\big)+\int\mathrm{d}^2\theta
\ W\big(\mathcal{Z}^A\big)+\int\mathrm{d}^2\theta^* \
\bar{W}\big(\mathcal{Z}_{\bar{A}}\big) \,
\end{equation}
with $W$ a holomorphic function of the non-associative algebra which
describes the interaction:
\begin{equation}
W=-\frac{\kappa_1}{8}\epsilon_{ABCD}\mathrm{Tr}\big(\mathcal{Z}^A,\big[\mathcal{Z}^B,\mathcal{Z}^C,\mathcal{Z}^D\big]\big)
\, .
\end{equation}
In fact, closure of the full $\mathcal{N}=8$ theory requires $\kappa_1=-\frac{1}{6}$\cite{Bagger:2007jr}.

\section{A Proposal for a Marginal Deformation \label{defsec}}

In \cite{Lunin:2005jy,Berman:2007tf}, the supergravity dual description of multiple membranes was studied
and possible deformations preserving the $AdS_4$ structure were
investigated. The preservation of the $AdS_4$ structure indicates the
possibility of deforming the membrane theory while preserving
conformal invariance. The deformation of the $S^7$ of course indicates
that the deformation of the theory wouldn't preserve the $SO(8)$
R-symmetry and hence the ${\cal{N}}=8$ supersymmetry. This is the
M-theory analogue of the Leigh Strassler deformation of ${\cal{N}}=4$
Yang-Mills whose supergravity dual was determined by Lunin and
Maldacenca \cite{Lunin:2005jy}.

We will proceed by describing properties of the Leigh Strassler
deformation and then describe the proposed M-theory analogue for the
membrane theory. The result will be to allow a marginal deformation
that preserves the ${\cal{N}}=2$ structure.

The Leigh Strassler deformation works by introducing a deformed
product in the field theory parametrised by $\beta$ (for this reason
it is also referred to as $\beta$ deformed Yang-Mills):
\begin{eqnarray}
f*g&=&e^{i\pi\beta\big(Q_f^1Q_g^2-Q_g^1Q_f^2\big)}f\cdot g\nonumber\\
&=&e^{i\pi\beta Q_{[f,}^1 Q_{g]}^2}f\cdot g \, ,  \label{def}
\end{eqnarray}
where $Q_f^i,Q_g^j$ denote the charges of the fields with respect to two
global U(1) symmetries. As such the deformation picks out two U(1)'s
as a special subgroup of the SO(6) R-symmetry.

This deformation only produces an effect within the
superpotential. Expressed in $\mathcal{N}=1$ superspace using chiral
fields $\Phi^I, \ I=1,2,3$ the superpotential is
\begin{equation}
W\sim \epsilon^{IJK}\Phi^I \Phi^J\Phi^K \, .
\end{equation}
After deforming this using (\ref{def}) the superpotential becomes
\begin{equation}
\Phi^1\big[\Phi^2,\Phi^3\big]_*=\Phi^1\big(e^{i\pi\beta}\Phi^2\Phi^3-e^{-i\pi\beta}\Phi^3\Phi^2\big)\, .
\end{equation}

Properties of the deformation can be summarised as follows;

\begin{enumerate}
	\item The deformed product relies on the use of two U(1)'s.
	\item The phase factor is defined in terms of the $U(1)_1\times U(1)_2$ 
charges of the fields and involves a commutator over the field indices.
	\item All products involving two different fields pick up a
              phase factor but effectively only
commutators of independent fields are modified by the deformation.
	\item The star product with any third field produces no new phases, for example
\begin{equation}
\Phi^1*\big(\Phi^2*\Phi^3\big)=\Phi^1\cdot\big(\Phi^2*\Phi^3\big)\, .
\end{equation}
\end{enumerate}

One should also note that the deformed product is very reminiscent of
the noncommutative Moyal product and indeed from the D-brane
perspective it is as if the transverse
space to the brane has been made noncommutative. This analogy becomes more explicit
from the perspective of the supergravity dual where one begins with
the nondeformed brane solutions and then carries out a
series of solution generating transformations to switch on background
fields. One can then make the choice of whether the background fields are on the
brane world volume, in which case the resulting deformation is to make the theory
noncommutative, or transverse to the brane world volume in which case one
produces this Leigh Strassler deformation.

We will now attempt now to generalise this deformation to the Bagger
and Lambert theory by introducing a deformed triple product with analogous properties to the deformed Lie bracket. We introduce a phase factor in the triple product which depends on the
charges of the fields under the $U(1)_1\times U(1)_2\times U(1)_3$
global symmetry. For chiral fields labelled $A,B,C$ we propose the following deformed associator
\begin{equation}
<A,B,C>_*=e^{i\pi\beta Q^1_{[A,}Q^2_{B,}Q^3_{C]}}<A,B,C>\, \label{3def} ,
\end{equation}
where the $Q^i$ denote the charges of the field under the
$U(1)_1\times U(1)_2\times U(1)_3$ global symmetry.
For all $A,B,C$ this gives the phase factor
\begin{eqnarray}
Q^1_{[A,}Q^2_{B,}Q^3_{C]}&=&Q^1_AQ^2_BQ^3_C+Q^1_BQ^2_CQ^3_A+Q^1_CQ^2_AQ^3_B\nonumber\\
&-&Q^1_CQ^2_BQ^3_A-Q^1_BQ^2_AQ^3_C-Q^1_AQ^2_CQ^3_B\\
&=&\mathrm{Det}\mathcal{Q}\, ,
\end{eqnarray}
where $\mathcal{Q}$ is the matrix of charges denoted by
\begin{equation}
\mathcal{Q}=\Bigg(\begin{array}{ccc}
  Q^1_A & Q^2_A & Q^3_A \\
  Q^1_B & Q^2_B & Q^3_B \\
  Q^1_C & Q^2_C & Q^3_C
\end{array}\Bigg).
\end{equation}

From this anti-symmetrisation it can be seen that cyclic permutations
of $<A,B,C>$ will deform with the same phase structure
as will cyclic permutations of $<C,B,A>$.
Furthermore, the multiplicative phase factor accompanying the latter
will be the reciprocal of the former.
This deformation of products of three fields results in
many nice features, which are
analogous to the Leigh Strassler deformation.
In particular, the three bracket of the nonassociative algebra deforms in a manner analogous to the
deformation of the commutator, that is:
\begin{eqnarray}
[\mathcal{Z}_A,\mathcal{Z}_B,\mathcal{Z}_C]_* &=& e^{i\pi\beta Q^1_{[A,}Q^2_{B,}Q^3_{C]}} \ \ \ \big(<\mathcal{Z}_A,\mathcal{Z}_B,\mathcal{Z}_C> \ + \ \mathrm{cyclic} \ \big)\nonumber\\
&-&e^{-i\pi\beta Q^1_{[A,}Q^2_{B,}Q^3_C]} \ \big(<\mathcal{Z}_C,\mathcal{Z}_B,\mathcal{Z}_A>  \ + \ \mathrm{cyclic} \ \big),
\end{eqnarray}
for all values of $A,B,C$. We find that properties of the deformation can be summarised as follows:

\begin{enumerate}
	\item The deformed product relies on the use of three U(1)'s.
	\item The phase factor is defined in terms of the $U(1)_1\times U(1)_2\times U(1)_3$ 
charges of the fields and involves a triple product over the field indices.
	\item All products involving three different fields pick up a
              phase factor but effectively only
triple products of independent fields are modified by the deformation.
	\item The star product with any fourth field produces no new phases.
\end{enumerate}

We find preservation of the algebraic properties of the
theory. Furthermore, the reality condition is respected by the deformed
triple product.
For the purpose of explicit computation, the chiral fields have been assigned the following charges
\begin{eqnarray}
\big(\mathcal{Z}^1,\mathcal{Z}^2,\mathcal{Z}^3,\mathcal{Z}^4\big)&\rightarrow & \big(\mathcal{Z}^1,\mathcal{Z}^2,e^{-i\varphi_1}\mathcal{Z}^3,e^{i\varphi_1}\mathcal{Z}^4\big): \ U(1)_1,\\
&\rightarrow & \big(\mathcal{Z}^1,e^{-i\varphi_2}\mathcal{Z}^2,e^{i\varphi_2}\mathcal{Z}^3,\mathcal{Z}^4\big): \ U(1)_2,\\
&\rightarrow & \big(e^{i\varphi_3}\mathcal{Z}^1,e^{-i\varphi_3}\mathcal{Z}^2,\mathcal{Z}^3,\mathcal{Z}^4\big): \ U(1)_3.
\end{eqnarray}
For all possible choices of $A\neq B\neq C\in(1,2,3,4)$ preserving the ordering in even permutations of $(1,2,3,4)$ the deformation is given by
\begin{eqnarray}
\big[\mathcal{Z}^A,\mathcal{Z}^B,\mathcal{Z}^C\big]_*&=&e^{i\pi\beta}\big(<\mathcal{Z}^A,\mathcal{Z}^B,\mathcal{Z}^C>+ \ \mathrm{cyclic} \ \big)\nonumber\\
&-&e^{-i\pi\beta}\big(<\mathcal{Z}^C,\mathcal{Z}^B,\mathcal{Z}^A>+ \ \mathrm{cyclic} \ \big),
\end{eqnarray}
provided the sequence $(A,B,C,D)$ in
\begin{equation}
\mathrm{Tr}\big(\mathcal{Z}^A,\big[\mathcal{Z}^B,\mathcal{Z}^C,\mathcal{Z}^D\big]_*\big),
\end{equation}
can be written as a positive permutation of $(1,2,3,4)$.
The cyclicity of the trace and the presence of a totally antisymmetric
tensor in the superpotential ensure this can be
done for every term in the action.
This ordering is important for re-expressing the deformation simply.
It can then be seen that the deformation preserves the properties of the associator within the trace
\begin{equation}
\mathrm{Tr}\big(\mathcal{Z}^A,<\mathcal{Z}^B,\mathcal{Z}^C,\mathcal{Z}^D>_*\big)=\mathrm{Tr}\big(<\mathcal{Z}^B,\mathcal{Z}^C,Z^D>_*,\mathcal{Z}^A\big)\, ,
\end{equation}
for all inequivalent $A,B,C,D \in (1,2,3,4)$ and that using this property we can then show
\begin{equation}
\mathrm{Tr}\big(\mathcal{Z}^A,\big<\mathcal{Z}^B, \mathcal{Z}^C,\mathcal{Z}^D\big>_*\big)=-\mathrm{Tr}\big(\big<\mathcal{Z}^A,\mathcal{Z}^B,\mathcal{Z}^C\big>_*,\mathcal{Z}^D\big)\, .
\end{equation}
It then follows for the triple product
\begin{equation}
\mathrm{Tr}\big(\mathcal{Z}^A,\big[\mathcal{Z}^B,\mathcal{Z}^C,\mathcal{Z}^D\big]_*\big)=-\mathrm{Tr}\big(\big[\mathcal{Z}^A,\mathcal{Z}^B,\mathcal{Z}^C\big]_*,\mathcal{Z}^D\big)\, .
\end{equation}
Analysis of anti-chiral fields reveals the same results with all
exponential factors mapping to their reciprocal. Also note that,
triple products containing triple products obey:
\begin{equation}
\big[\alpha,\beta,\big[A,B,C\big]_*\big]_* =
\big[\alpha,\beta,\big[A,B,C\big]_*]  \, .
\end{equation}

As for $\mathcal{N}=4$ super Yang-Mills, the effects of the deformation on the action are found only in the superpotential. The $\mathcal{N}=2$ Lagrangian becomes
\begin{equation}
\mathcal{L}_*=\frac{1}{2}\int\mathrm{d}^4\theta \ \mathrm{Tr}\big(\mathcal{Z}^A,\mathcal{Z}_{\bar{A}}\big) - \frac{\kappa_1}{8}\epsilon_{ABCD}\int\mathrm{d}^2\theta\mathrm{Tr}\big(\mathcal{Z}^A,\big[\mathcal{Z}^B,\mathcal{Z}^C,\mathcal{Z}^C\big]_*\big) + \int\mathrm{d}^2\theta^*\bar{\mathcal{W}}_*(\mathcal{Z}_{\bar{A}}).
\end{equation}
The supersymmetry algebra remains unchanged along with the superspace constraint
\begin{equation}
\big[\mathcal{Z}^A,\mathcal{Z}^B,\mathcal{Z}_{\bar{B}}\big]=0\, .
\end{equation}
That the deformed theory can be expressed in $\mathcal{N}=2$ superspace means that $\mathcal{N}=2$ supersymmetry is manifest. In fact one may have preserved more but we expect only one quarter of the supersymmetry to be preserved as in Leigh Strassler.

Note that although we began with the $\mathcal{N}=2$ version of the
multi-membrane theory, we may consider this as a deformation to the
full $\mathcal{N}=8$ theory that preserves ${\mathcal{N}}=2$ supersymmetry. The route
via the $\mathcal{N}=2$ superspace version of the membrane was chosen because of familiarity with the $\beta$-deformation of
$\mathcal{N}=4$ SYM theory and as in that case is much easier to
formulate using the superfield language.

We would like to see the effect of the deformation on the bosonic scalar potential:
\begin{equation}
V=\frac{1}{2\cdot 3!} \ \mathrm{Tr}\big(\big[X^I,X^J,X^K\big],\big[X^I,X^J,X^K\big]\big).
\end{equation}
To start with we will look at the terms arising in the undeformed
superpotential that survive the Grassman integration.
There are four different types of terms that do this. These are $\big(Z,[Z,\psi,\psi]\big), \ \big(\psi,[Z,Z,\psi]\big), \ \big(F,[Z,Z,Z]\big) \ \mathrm{and} \ \big(Z,[Z,Z,F]\big)$. We neglect the fermions for the moment. Symmetries of the triple-product and the properties of the trace allow us to group terms
\begin{eqnarray}
\mathcal{W}|_{\theta\theta}&=&-\frac{\kappa_1}{8}\epsilon^{ABCD}\mathrm{Tr}\big(Z^A,\big[Z^B,Z^C,\epsilon^{DEFG}\big[Z_{\bar{E}},Z_{\bar{F}},Z_{\bar{G}}\big]\big]\big)\nonumber\\
& & -\frac{\kappa_1}{8}\epsilon^{ABCD}\mathrm{Tr}\big(Z^A,\big[Z^B,\epsilon^{CEFG}\big[Z_{\bar{E}},Z_{\bar{F}},Z_{\bar{G}}\big],Z^D\big]\big)\nonumber\\
& &-\frac{\kappa_1}{8}\epsilon^{ABCD}\mathrm{Tr}\big(Z^A,\big[\epsilon^{BEFG}\big[Z_{\bar{E}},Z_{\bar{F}},Z_{\bar{G}}\big],Z^C,Z^D\big]\big)\nonumber\\
& &-\frac{\kappa_1}{8}\epsilon^{ABCD}\mathrm{Tr}\big(\epsilon^{AEFG}\big[Z_{\bar{E}},Z_{\bar{F}},Z_{\bar{G}}\big],Z^B,Z^C,Z^D\big]\big),
\end{eqnarray}
which becomes
\begin{eqnarray}
\mathcal{W}|_{\theta\theta}&=&-\frac{24\kappa_1}{8}\mathrm{Tr}\big(Z^A,\big[\big[Z^C,Z^D,Z_{\bar{C}}\big],Z_{\bar{D}},Z_{\bar{A}}\big]\big)\nonumber\\
& &-\frac{24\kappa_1}{8}\mathrm{Tr}\big(Z^A,\big[Z_{\bar{C}},\big[Z^C,Z^D,Z_{\bar{D}}\big],Z_{\bar{A}}\big]\big)\nonumber\\
& &-\frac{24\kappa_1}{8}\mathrm{Tr}\big(Z^A,\big[Z_{\bar{C}},Z_{\bar{D}},\big[Z^C,Z^D,Z_{\bar{A}}\big]\big]\big).
\end{eqnarray}
The first and second terms vanish using the constraint
\begin{equation}
\big[Z^A,Z^B,Z_{\bar{B}}\big]=0,
\end{equation}
whilst, using the symmetries of the trace and triple-product gives
\begin{eqnarray}
\mathcal{W}|_{\theta\theta}&=&-\frac{24\kappa_1}{8}\mathrm{Tr}\big(Z^A,\big[Z_{\bar{C}},Z_{\bar{D}},\big[Z^C,Z^D,Z_{\bar{A}}\big]\big]\big)\nonumber\\
&=&-\frac{24\kappa_1}{8}\mathrm{Tr}\bigg(Z^A,\Big(\big[\big[Z_{\bar{C}},Z_{D},Z^C\big],Z^D,Z_{\bar{A}}\big]\nonumber\\
& &\ \ \ \ \ \ \ \ \ \ \ \ \ \ \ \ \ \ \ +\big[Z^C,\big[Z_{\bar{C}},Z_{\bar{D}},Z^D\big],Z_{\bar{A}}\big]\nonumber\\
& &\ \ \ \ \ \ \ \ \ \ \ \ \ \ \ \ \ \ \ +\big[Z^C,Z^D,\big[Z_{\bar{C}},Z_{\bar{D}},Z_{\bar{A}}\big]\big]\Big)\bigg),\nonumber\\
&=&-\frac{24\kappa_1}{8}\mathrm{Tr}\big(Z^A,\big[Z^C,Z^D,\big[Z_{\bar{C}},Z_{\bar{D}},Z_{\bar{A}}\big]\big]\big),\nonumber\\
&=&\frac{24\kappa_1}{8}\mathrm{Tr}\big(\big[Z^A,Z^B,Z^C],\big[Z_{\bar{A}},Z_{\bar{B}},Z_{\bar{C}}\big]\big).
\end{eqnarray}
In the undeformed theory, this term is proportional to the term quadratic in the auxiliary field, which arises from the chiral superfield kinetic term. These combine to produce the sixth order scalar field term in the potential of the component Lagrangian.

The chiral kinetic term of the $\mathcal{N}=2$ superfield Lagrangian is invariant under the deformation but we have already seen that the triple-products, present in the superpotential, are modified while the constraint equations are not. The field equation for the auxiliary field becomes
\begin{equation}
F^A=\kappa_1\epsilon^{ABCD}\big[Z_{\bar{B}},Z_{\bar{C}},Z_{\bar{D}}\big]_*\, .
\end{equation}
After the deformation, the superpotential can be written, using the deformed symmetries of the trace and triple product, as
\begin{equation}
\mathcal{W}_*=-3\kappa_1\mathrm{Tr}\big(\mathcal{Z}^1,\big[\mathcal{Z}^2,\mathcal{Z}^3,\mathcal{Z}^4\big]_*\big)\, .
\end{equation}
That both the auxiliary field equation and the superpotential can be
simply expressed in the deformed theory by replacing the triple
product with a star-triple product is due to the symmetries of the
star product outlined earlier.

As for the undeformed case, the only purely bosonic contribution coming from the superpotential is of the same form as the term coming from the chiral kinetic piece and is given by
\begin{equation}
\mathcal{W}|_{\theta\theta}\sim \mathrm{Tr}\big(\big[Z^A,Z^B,Z^C\big]_*,\big[Z_{\bar{A}},Z_{\bar{B}},Z_{\bar{C}}\big]_*\big)\, .
\end{equation}

\subsection{The $\beta$-Deformation related to the supergravity dual}

We now wish to examine other deformations, which may be related to deformations of the $AdS_4\times S^7$ supergravity dual of the membrane. Matching such deformations will provide further support to the Bagger Lambert conjectured multi-membrane theory.

The deformations in \cite{Berman:2007tf} preserve
the $AdS_4$ (and hence the conformality of the membrane theory) but deform the $S^7$ using M-theory solution generating
transformations. These solution generating transformations involve
identifying a three torus and then acting with a solution generating
transformation on a $T^2 \subset T^3$ to deform the solution.
It was shown that different choices of the $T^2$  produced distinct
solutions. 
A whole spectrum of solutions could be generated with a great
diversity in their apparent properties. 
In one case the effects
of the deformation were such that the entropy of
related black-brane was left invariant by the deformation. In
another case, the deformations produced a different entropy involving a simple multiplicative
factor with the deformation parameter. There were also more
complicated examples.

Motivated by the deformation of the supergravity dual, of a $T^2$, we
propose the deformation of the membrane theory should be with a
deformed two product. The deformed associator, given by replacing the
regular product with the star product as given in (\ref{def}), is

\begin{equation}
<A,B,C>_*=A*(B*C)-(A*B)*C\, .
\end{equation}

For star products of the nonassociative theory we have no reason to
assume that we will recover the properties that were found for the star product in $\mathcal{N}=4$ SYM or those described in section 5. 
Even if these properties were preserved then 
we could not necessarily expect the two terms present in the
associator 
to transform in an homogenous manner. It would then be impossible to express the deformed associator as
a product of a phase 
(containing all the information of the deformation) and the undeformed
associator as we did in (\ref{3def}). Such a factorisation was crucial in allowing us to express the deformed action simply. 

Interestingly, for some choices of $T^2$ this factorisation property is
present. For other choices we lose this 
property and star-products with
third fields 
produce additional phase-factors under the deformation. 
However, this can occur in such a way as to ensure that the two
constituent terms in 
any associator transform in an homogenous manner. Again, we find we
are able to 
generate families of solutions through the deformation with very
different properties. 
All results depend on the particular choice of fields and charge assignments of the $U(1)\times U(1)$ subgroup.
Let us see how the associator deforms under the star product. Using this deformation, the product of three fields always transforms like

\begin{equation}
A*(B*C)=e^{i\pi\beta Q^1_{[A,}Q^2_{BC]}}e^{i\pi\beta Q^1_{[B,}Q^2_{C]}} \ A\cdot(B\cdot C)\, ,
\end{equation}
where $Q^i_{BC}$ is the charge of $(B\cdot C)$ under the global
symmetry $U(1)_i$. Additivity of charge for this global symmetry gives
the phase as
\begin{equation}
e^{i\pi\beta \big(Q^1_{[A,}Q^2_{B]}+Q^1_{[A,}Q^2_{C]}+Q^1_{[B,}Q^2_{C]}\big)},
\end{equation}
where the ordering of the field indices is determined by the ordering
of the fields in the product. 
We can see now, that given three fields the constituent terms within
the associator 
can be expected to transform in the same way under the deformation. 
Furthermore, we can conclude that associators related by complete anti-symmetry
\begin{equation}
<A,B,C> \ \longleftrightarrow \ <C,B,A>
\end{equation}
will pick up reciprocal phase factors. However we lose the ability to
relate phase factors for cyclic 
permutations of all associators. This is a result of the replacement
of a product of three field charges, 
anti-symmetrised on the field indices, with three commutators of field
charges. This symmetry is lost for deformations of general associators
and triple products under arbitrary global $U(1)\times U(1)$'s. 

We look now at two particular examples to highlight the behaviour of
the non-associative Bagger Lambert theory under this deformation.

\subsection{Example I:$\big[\mathcal{Z}^2,\mathcal{Z}^3,\mathcal{Z}^4\big]$ under $U(1)_1\times U(1)_3$ }

In this example, the star-product with a third field produces a new phase factor under the deformation. This happens in such a way as to ensure each associator transforms into the old undeformed associator and a multiplicative phase factor. Using the arbitrary charge assignments from before and deforming under $U(1)_1\times U(1)_3$ we see

\begin{eqnarray}
\big[\mathcal{Z}^2,\mathcal{Z}^3,\mathcal{Z}^4\big]_*&=& <\mathcal{Z}^2,\mathcal{Z}^3,\mathcal{Z}^4>-<\mathcal{Z}^4,\mathcal{Z}^3,\mathcal{Z}^2>\nonumber\\
&+&<\mathcal{Z}^3,\mathcal{Z}^4,\mathcal{Z}^2>-<\mathcal{Z}^2,\mathcal{Z}^4,\mathcal{Z}^3>\nonumber\\
&+&e^{-2\pi i\beta} \ <\mathcal{Z}^4,\mathcal{Z}^2,\mathcal{Z}^3>-e^{2\pi i\beta}<\mathcal{Z}^3,\mathcal{Z}^2\mathcal{Z}^4>.
\end{eqnarray} 
The triple products do not transform in a similar manner to that of the $U(1)\times U(1)\times U(1)$ deformation defined earlier. However, it may be relatable to one of the less symmetric cases found for the M2 supergravity dual.

\subsection{Example II:$\big[\mathcal{Z}^1,\mathcal{Z}^2,\mathcal{Z}^3\big]$ under $U(1)_2\times U(1)_3$ }

In this example, we find that star-products with a third field produce
no new deformative phase factors just as we found in with the original
$\beta$-deformation. The deformation can therefore be fully expressed 
in terms of a deformation of the first product taken within any
associator. 
Furthermore, for this particular choice, each and every bi-linear
star-product produces exactly the correct phase factor to reproduce 
the effects of the deformation defined using three global $U(1)$
symmetries. For two deformations with very different origins it is 
surprising that we find the same resultant deformation of
triple-product 
terms within the superpotential. We find, under $U(1)_2\times U(1)_3$,

\begin{eqnarray}
\big[\mathcal{Z}^1,\mathcal{Z}^2,\mathcal{Z}^3\big]_*&=&<\mathcal{Z}^1,\mathcal{Z}^2,\mathcal{Z}^3>_*+<\mathcal{Z}^2,\mathcal{Z}^3,\mathcal{Z}^1>_*\nonumber\\
&+&<\mathcal{Z}^3,\mathcal{Z}^1,\mathcal{Z}^2>_*-<\mathcal{Z}^3,\mathcal{Z}^2,\mathcal{Z}^1>_*\nonumber\\
&-&<\mathcal{Z}^2,\mathcal{Z}^1,\mathcal{Z}^3>_*-<\mathcal{Z}^1,\mathcal{Z}^3,\mathcal{Z}^2>_*.
\end{eqnarray}
On further investigation it was found that third and higher order star products were irrelevant in the deformation
\begin{eqnarray}
\big[\mathcal{Z}^1,\mathcal{Z}^2,\mathcal{Z}^3\big]_*&=&\mathcal{Z}^1\big(\mathcal{Z}^2*\mathcal{Z}^3\big)-\big(\mathcal{Z}^1*\mathcal{Z}^2\big)\mathcal{Z}^3+\mathcal{Z}^2\big(\mathcal{Z}^3*\mathcal{Z}^1\big)\nonumber\\
&-&\big(\mathcal{Z}^2*\mathcal{Z}^3\big)\mathcal{Z}^1+\mathcal{Z}^3\big(\mathcal{Z}^1*\mathcal{Z}^2\big)-\big(\mathcal{Z}^3*\mathcal{Z}^1\big)\mathcal{Z}^2\nonumber\\
&-&\mathcal{Z}^3\big(\mathcal{Z}^2*\mathcal{Z}^2\big)+\big(\mathcal{Z}^3*\mathcal{Z}^2\big)\mathcal{Z}^1-\mathcal{Z}^2\big(\mathcal{Z}^1*\mathcal{Z}^3\big)\nonumber\\
&+&\big(\mathcal{Z}^2*\mathcal{Z}^1\big)\mathcal{Z}^3-\mathcal{Z}^1\big(\mathcal{Z}^3*\mathcal{Z}^2\big)+\big(\mathcal{Z}^1*\mathcal{Z}^3\big)\mathcal{Z}^2.
\end{eqnarray} 
This can be written
\begin{eqnarray}
\big[\mathcal{Z}^1,\mathcal{Z}^2,\mathcal{Z}^3\big]_*&=&e^{i\pi \beta}\big(<\mathcal{Z}^1,\mathcal{Z}^2,\mathcal{Z}^3>+\mathrm{cyclic}\big)\nonumber\\
&-&e^{-i\pi\beta} \big(<\mathcal{Z}^3,\mathcal{Z}^2,\mathcal{Z}^1> + \
\mathrm{cyclic}\big) \, .
\end{eqnarray}
Furthermore, if we then take the product with $\mathcal{Z}^4$ from 
\begin{equation}
\mathrm{Tr}\Big(\mathcal{Z}^4,\big[\mathcal{Z}^1,\mathcal{Z}^2,\mathcal{Z}^3\big]_*\Big)_*
\end{equation}
we obtain 
\begin{equation}
\mathrm{Tr}\Big(\mathcal{Z}^4,\big[\mathcal{Z}^1,\mathcal{Z}^2,\mathcal{Z}^3\big]_*\big),
\end{equation}
which produces exactly the same term that we would generate using the
$U(1)_1\times U(1)_2\times U(1)_3$ deformation completely
anti-symmetrised over the three field indices as already described in
section \ref{defsec}. Thus the marginal deformation described in
section \ref{defsec} will be dual to the deformation described in
\cite{Berman:2007tf}.

\section{Conclusions}

We are still some way from a full understanding the interacting membrane
theory but there are significant indications that we are heading in
the right direction. One, we have reproduced aspects of the fivebrane
from the open membrane. Two, we have explicit examples of
supersymmetry preserving marginal deformations that may be related to
deformations of the supergravity dual. The key issue that needs to be further explored is to understand
how the number of membranes enters the theory and of course relate
this to the number of degrees of freedom and hopefully to the proposal described in
\cite{Berman:2006eu}.

An immediate technical
question is the study of the membrane supersymmetry in the presence of
a boundary and the supersymmetry of the boundary theory. One can also
try to understand quantum properties of the twisted Chern-Simons
theory. Famously, the partition function of Chern-Simons theory
\cite{Witten:1988hf} leads
to the Ray-Singer Torsion of the three manifold. The role of
the Ray-Singer torsion of the membrane and the partition function of
the Chern-Simons theory at level -1 (coming from the anti-self-dual
sector) are still very much open questions.

\bigskip
{\bf{Note Added}}: 

As this paper was being prepared, two interesting preprints appeared
discussing aspects of the Bagger and Lambert theory \cite{today1,today2}.

\section{Acknowledgements}

This work was in part supported by the EC Marie Curie
Research Training Network, MRTN-CT-2004-512194.
LT and DT are supported by an STFC (nee PPARC) grant.
We are gratefull to Tom Brown, Paul Heslop, Sanjaye Ramgoolam and Bill Spence for
discussions.


\begin{thebibliography}{99}

%\cite{Berman:2007bv}
\bibitem{Berman:2007bv}
  D.~S.~Berman,
  ``M-theory branes and their interactions,''
  Phys.\ Rept.\  {\bf 456} (2008) 89
  [arXiv:0710.1707 [hep-th]].
  %%CITATION = PRPLC,456,89;%%

%\cite{Bagger:2006sk}
\bibitem{Bagger:2006sk}
  J.~Bagger and N.~Lambert,
  ``Modeling multiple M2's,''
  Phys.\ Rev.\  D {\bf 75} (2007) 045020
  [arXiv:hep-th/0611108].
  %%CITATION = PHRVA,D75,045020;%%


%\cite{Bagger:2007jr}
\bibitem{Bagger:2007jr}
  J.~Bagger and N.~Lambert,
  ``Gauge Symmetry and Supersymmetry of Multiple M2-Branes,''
  arXiv:0711.0955 [hep-th].
  %%CITATION = ARXIV:0711.0955;%%

%\cite{Bagger:2007vi}
\bibitem{Bagger:2007vi}
  J.~Bagger and N.~Lambert,
  ``Comments On Multiple M2-branes,''
  JHEP {\bf 0802} (2008) 105
  [arXiv:0712.3738 [hep-th]].
  %%CITATION = JHEPA,0802,105;%%

%\cite{Gustavsson:2007vu}
\bibitem{Gustavsson:2007vu}
  A.~Gustavsson,
  ``Algebraic structures on parallel M2-branes,''
  arXiv:0709.1260 [hep-th].
  %%CITATION = ARXIV:0709.1260;%%

  %\cite{Berman:2001fs}
\bibitem{Berman:2001fs}
  D.~S.~Berman and P.~Sundell,
  ``AdS(3) OM theory and the self-dual string or membranes ending on the
  five-brane,''
  Phys.\ Lett.\  B {\bf 529}, 171 (2002)
  [arXiv:hep-th/0105288].
  %%CITATION = PHLTA,B529,171;%%

%\cite{Lunin:2005jy}
\bibitem{Lunin:2005jy}
  O.~Lunin and J.~M.~Maldacena,
  ``Deforming field theories with U(1) x U(1) global symmetry and their
  gravity duals,''
  JHEP {\bf 0505}, 033 (2005)
  [arXiv:hep-th/0502086].
  %%CITATION = JHEPA,0505,033;%%

%\cite{Berman:2007tf}
\bibitem{Berman:2007tf}
  D.~S.~Berman and L.~C.~Tadrowski,
  ``M-Theory Brane Deformations,''
  Nucl.\ Phys.\  B {\bf 795}, 201 (2008)
  [arXiv:0709.3059 [hep-th]].
  %%CITATION = NUPHA,B795,201;%%
%\cite{Berman:2002kd}
\bibitem{Berman:2002kd}
  D.~S.~Berman and E.~Rabinovici,
  ``Supersymmetric gauge theories,''
  arXiv:hep-th/0210044.
  %%CITATION = HEP-TH/0210044;%%

%\cite{Howe:1997ue}
\bibitem{Howe:1997ue}
  P.~S.~Howe, N.~D.~Lambert and P.~C.~West,
  ``The self-dual string soliton,''
  Nucl.\ Phys.\  B {\bf 515}, 203 (1998)
  [arXiv:hep-th/9709014].
  %%CITATION = NUPHA,B515,203;%%
%\cite{Strominger:1995ac}

\bibitem{Strominger:1995ac}
  A.~Strominger,
  ``Open p-branes,''
  Phys.\ Lett.\  B {\bf 383}, 44 (1996)
  [arXiv:hep-th/9512059].
  %%CITATION = PHLTA,B383,44;%%

%\cite{Townsend:1995af}
\bibitem{Townsend:1995af}
  P.~K.~Townsend,
  ``D-branes from M-branes,''
  Phys.\ Lett.\  B {\bf 373} (1996) 68
  [arXiv:hep-th/9512062].
  %%CITATION = PHLTA,B373,68;%%

%\cite{Gustavsson:2008dy}
\bibitem{Gustavsson:2008dy}
  A.~Gustavsson,
  ``Selfdual strings and loop space Nahm equations,''
  arXiv:0802.3456 [hep-th].
  %%CITATION = ARXIV:0802.3456;%%

%\cite{Kawamura:2003cw}
\bibitem{Kawamura:2003cw}
  Y.~Kawamura,
  ``Cubic matrix, generalized spin algebra and uncertainty relation,''
  Prog.\ Theor.\ Phys.\  {\bf 110} (2003) 579
  [arXiv:hep-th/0304149].
  %%CITATION = PTPKA,110,579;%%

%\cite{tHooft:1976fv}
\bibitem{tHooft:1976fv}
  G.~'t Hooft,
  ``Computation of the quantum effects due to a four-dimensional
  pseudoparticle,''
  Phys.\ Rev.\  D {\bf 14}, 3432 (1976)
  [Erratum-ibid.\  D {\bf 18}, 2199 (1978)].
  %%CITATION = PHRVA,D14,3432;%%


  %\cite{Witten:1983ar}
\bibitem{Witten:1983ar}
  E.~Witten,
  ``Nonabelian bosonization in two dimensions,''
  Commun.\ Math.\ Phys.\  {\bf 92}, 455 (1984).
  %%CITATION = CMPHA,92,455;%%

  %\cite{Witten:1988hf}
\bibitem{Witten:1988hf}
  E.~Witten,
  ``Quantum field theory and the Jones polynomial,''
  Commun.\ Math.\ Phys.\  {\bf 121}, 351 (1989).
  %%CITATION = CMPHA,121,351;%%

  %\cite{Elitzur:1989nr}
\bibitem{Elitzur:1989nr}
  S.~Elitzur, G.~W.~Moore, A.~Schwimmer and N.~Seiberg,
  ``Remarks On The Canonical Quantization Of The Chern-Simons-Witten Theory,''
  Nucl.\ Phys.\  B {\bf 326}, 108 (1989).
  %%CITATION = NUPHA,B326,108;%%

%\cite{Dunne:1998qy}
\bibitem{Dunne:1998qy}
  G.~V.~Dunne,
  ``Aspects of Chern-Simons theory,''
  arXiv:hep-th/9902115.
  %%CITATION = HEP-TH/9902115;%%

%\cite{Berman:2006eu}
\bibitem{Berman:2006eu}
  D.~S.~Berman and N.~B.~Copland,
  ``A note on the M2-M5 brane system and fuzzy spheres,''
  Phys.\ Lett.\  B {\bf 639} (2006) 553
  [arXiv:hep-th/0605086].
  %%CITATION = PHLTA,B639,553;%%

\bibitem{today1}
S.~Mukhi and C.~Papageorgakis,
  ``M2 to D2,''
  arXiv:0803.3218 [hep-th].
  %%CITATION = ARXIV:0803.3218;%%
\bibitem{today2}
M.~A.~Bandres, A.~E.~Lipstein and J.~H.~Schwarz,
  ``N = 8 Superconformal Chern--Simons Theories,''
  arXiv:0803.3242 [hep-th].
  %%CITATION = ARXIV:0803.3242;%%



\end{thebibliography}
\end{document}